\documentclass[aps, 
               prb, 
               twocolumn, 
               10pt, 
               nofootinbib,
               superscriptaddress,
               amsfonts, 
               amsmath, 
               amssymb]{revtex4-2}

\usepackage{lmodern}
\usepackage{graphicx}
\usepackage[unicode,
            colorlinks=true,
            linkcolor=black,
            citecolor=black,
            urlcolor=black]{hyperref}

\usepackage{bm}
\usepackage{xcolor}
\usepackage{physics}
\usepackage{siunitx}
\DeclareMathOperator*{\argmax}{argmax}

\newcommand{\QDev}[0]{Center for Quantum Devices, Niels Bohr Institute, University of Copenhagen, DK-2100 Copenhagen, Denmark} 
\newcommand{\NanoLund}[0]{
Division of Solid State Physics and NanoLund, Lund University, S-22100 Lund, Sweden}

\begin{document}
\title{Nonsinusoidal current-phase relations in\\ semiconductor-superconductor-ferromagnetic insulator devices}
\author{Andrea Maiani}
\affiliation{\QDev}

\author{Karsten Flensberg}
\affiliation{\QDev}

\author{Martin Leijnse}
\affiliation{\QDev}
\affiliation{\NanoLund}

\author{Constantin~Schrade}
\affiliation{\QDev}

\author{Saulius Vaitiek\.enas}
\affiliation{\QDev}

\author{Rub\'en Seoane Souto}
\affiliation{\QDev}
\affiliation{\NanoLund}

\date{\today}

\begin{abstract}
Coherent tunneling processes of multiple Cooper pairs across a Josephson junction give rise to high harmonics in the current phase relation. In this work, we propose and study Josephson junctions based on semiconductor-superconductor-ferromagnetic insulator heterostructures to engineer nonsinusoidal current-phase relations. The gate-tunability of charge carriers density in the semiconductor, together with the adjustable magnetization of the ferromagnetic insulator, provides control over the content of the supercurrent harmonics. At finite exchange field, hybrid junctions can undergo a 0\,--\,$\pi$ phase transition, resulting in a supercurrent reversal. Close to the transition, single-pair tunneling is suppressed and the current-phase relation is dominated by the second-harmonic, indicating transport primarily by pairs of Cooper pairs. Finally, we demonstrate that non-collinear magnetization or spin-orbit coupling in the leads and the junction can lead to a gate-tunable Josephson diode effect with efficiencies of up to $\sim30\%$.
\end{abstract}
\maketitle

In Josephson junctions with insulating weak links, the flow of a dissipationless supercurrent arises from individual Cooper-pair tunneling events, which is typically characterized by a sinusoidal current-phase relation (CPR)~\cite{Golubov_RMP_2004}. For junctions with high transparency, additional contributions to the supercurrent appear thanks to the simultaneous coherent tunneling of multiple Cooper pairs~\cite{Beenakker_PRL_1991b, Bagwell_PRB_1992, Sochnikov_PRL2015, english2016observation, nanda2017current, Spanton_NatPhys2017, schindler2018higher, Kayyalha_NPJQM2020, nichele2020relating}. These tunneling events give rise to higher harmonics in the CPR, leading to deviations from the standard sinusoidal form. 

In a simple weak link, the Josephson energy is minimized when the superconducting phases on both sides of the junction are equal. However, when time-reversal symmetry is broken, a phase transition can occur, resulting in an equilibrium state with the relative phase difference of $\pi$. This shift in the macroscopic degree of freedom leads to a phase transition between the so-called $0$ and $\pi$ phases~\cite{Bulaevskii_JEPT_1977, Rozhkov_PRB_2001, VanDam_Nature_2006, Schrade_PRL_2015, Delagrange_PRB_2016}.

Josephson junctions with broken time-reversal symmetry can be tuned to a regime where both the $0$ and the $\pi$ phases are { local minima} of the Josephson potential~\cite{Radovic_PRB_2001, Sellier_PRL_2004}. In this case, the junction is said to be $0'$ or $\pi'$ regime if the global minimum is at $0$ or $\pi$ phase difference, { while the other state is a metastable phase}. The fundamental harmonic changes sign when moving from $0'$ to $\pi'$ and vanishes at the crossover between the two. At this point, the current is dominated by higher harmonics, causing a nonsinusoidal CPR~\cite{Houzet_PRB_2005}. { Other mechanisms for nonsinusoidal CPRs can be found in out-of-equilibrium systems, like resonator-induced deviations in a Zeeman split double quantum dot junction~\cite{Hussein_PRB_2021} and deviations induced by spin injection~\cite{Rezaei_PRR_2020}}

The study of such nonsinusoidal CPRs has recently gained significant interest due to potential applications in developing protected superconducting qubits~\cite{Smith_npjQI_2020, Larsen_PRL_2020, Guo_PRB_2022, Schrade_PRXQ_2022, Maiani_PRXQ_2022} and supercurrent diodes~\cite{Silaev_JPCM2014, Yokoyama_PRB2014, Halterman_PRB2022, Ilic_PRL2022, Ilic_PRAp2022, Davydova_SciAdv2022, Souto_PRL_2022, Mazur_arXiv2022,soori2023anomalous,steiner2022diode,tanaka2022theory,kokkeler2022field,haenel2022superconducting,fu2022gate, Trahms_2023,turini2022josephson,daido2022superconducting,lu2022josephson,wei2022supercurrent,fominov2022asymmetric,chiles2022non,legg2022superconducting,wang2022symmetry,song2022interference,gupta2022superconducting, Soori_2023, Ortega_Taberner_2023, Hu_diodeArxiv,legg2023parity}. These novel applications of Josephson junctions require not only a static nonsinusoidal CPR, but also the ability to control its harmonic content.

\begin{figure}[ht]
    \centering
    \includegraphics[width=1.0\columnwidth]{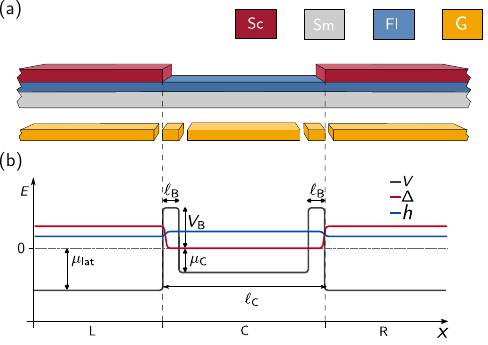}
    \caption{
    \textbf{Josephson junction in a semiconductor-ferromagnet-superconductor device.}
    Sketch of a ferromagnetic hybrid junction (a) and the {one-dimensional} continuum model considered in this paper (b). 
    The {semiconductor (Sm)} nanowire divides into three regions: the lateral left ($\mathrm{L}$) and right ($\mathrm{R}$) regions feature proximity-induced superconductivity from the superconductor {(Sc)} shells, while in the central ($\mathrm{C}$) region the superconductor is etched. { All the regions are subjected to the ferromagnetic proximity effect due to the ferromagnetic insulator inset (FI)}. The electron density in the central region and the barriers between the central and the lateral ones can be controlled by gates {(G)}.
    }
\label{fig:sketch}
\end{figure}

In this work, we propose and study a platform for tunable nonsinusoidal CPRs based on Josephson junctions comprised of semiconductor-superconductor-ferromagnetic insulator materials, Fig.~\ref{fig:sketch}. Recently, such platforms have been realized experimentally in InAs nanowires coated with epitaxial Al and EuS shells~\cite{Liu_NL_2019}, which exhibited signatures of spin-polarized subgap states~\cite{Vaitiekenas_NatPhys_2021,Vaitiekenas_PRB_2022}, and supercurrent reversal~\cite{Razmadze_2023}. We show that these hybrid materials  offer a new way to control the harmonic content of the CPR by combining the gate-tunable charge carriers density of the semiconductor and the adjustable magnetization of ferromagnetic insulator insets. Additionally, in these devices, both the superconducting pairing and the exchange field are induced in the semiconductor through proximity effects. This allows for unique regimes where the exchange field in the superconductor is below the Chandrasekhar-Clogston limit~\cite{Chandrasekhar_APL_1962, Clogston_PRL_1962}, while in the semiconductor it can exceed the induced pairing potential. As a result, this platform is suited for studying superconductivity under extreme exchange fields that can surpass the induced pairing potential.

To illustrate the prospects of this platform, we discuss an approach for realizing a pure second-harmonic CPR characterized by the coherent tunneling of pairs of Cooper pairs. We find this condition to be present both in the open and in the quantum dot regime.
We show that non-collinear magnetization, or, alternatively, non-collinear spin-orbit coupling fields in the leads and the junction allow for a gate-tunable Josephson diode effect with efficiencies up to $\sim30\%$. Our work expands on the possible technological applications of ferromagnetic hybrid devices beyond the realization of topological superconductivity by exploring the tunability of nonsinusoidal CPR in hybrid Josephson junctions. 

\section{Model}
The system we are considering can be conceptually split into three regions: two lateral ($\mathrm{L}$ and $\mathrm{R}$), and a central region ($\mathrm{C}$), see Fig.~\ref{fig:sketch}(a). The superconductor primarily induces a superconducting pairing potential $\Delta(x)$ in the semiconductor through the proximity effect. It also contributes to the electrostatic potential landscape $V(x) = - \mu(x)$. The ferromagnetic insulator induces an exchange field $\bm{h}(x)$ in both the superconductor and the semiconductor.
Given that both the exchange field and the pairing potential are induced in a semiconductor with controllable charge carrier density, there is not a fixed hierarchy of energy scales, and, in principle, all regimes can be realized in the system. We consider that the exchange field is sufficiently weak in the superconductor such that superconductivity persists. This condition is relaxed in the semiconductor, where the induced exchange field can overcome the induced pairing potential. Therefore, ferromagnetic hybrid junctions allow exploring the parameter space beyond the conventional regime ($\mu \gg \Delta > |\bm{h}|$). Note that we do not refer to a particular arrangement of the layers in the lateral region, as different combinations of interfaces, for instance, tripartite arrangement~\cite{Escribano_PRB_2021, Liu_PRB_2021} and tunneling arrangement~\cite{Maiani_PRB_2020, Escribano_NPJQM_2022}, allows induction of both superconducting pairing and exchange field in the semiconductor.

The system Hamiltonian is $H=\frac{1}{2}\int \psi^\dag \mathcal{H} \psi$, where the Bogoliubov-de Gennes (BdG) Hamiltonian $\mathcal{H}$ in the Nambu spinor basis $\psi^\dag = \begin{pmatrix}
\psi_\uparrow^\dag  & \psi_\downarrow^\dag  & -\psi_\downarrow  & \psi_\uparrow \end{pmatrix}$ is
\begin{equation}
\begin{split}
    \mathcal{H} = \Big[\frac{\hbar^2 k_x}{2 m^*} - \mu\Big]\tau_z + 
    \bm{h}\cdot \bm{\sigma} + \Delta \tau_+ + \Delta^\dag \tau_- + \mathcal{H}_\mathrm{SOC}\,.
\end{split}
\end{equation}
Here,  $k_x =-i\partial_x$ is the momentum operator (we consider a single mode in the junction), $m^*$ is the effective mass, and $\sigma_j$ and $\tau_j$ are the Pauli matrices in the spin and particle-hole space, respectively. The spin-orbit coupling Hamiltonian $\mathcal{H}_\mathrm{SOC}$ is given in Eq.~\eqref{eq:soc_hamiltonian} and discussed in Sec.~\ref{sec:soc}.

The proximity-induced exchange field $\bm{h}$ is due to the coupling to the ferromagnetic insulator and, in principle, can be spatially inhomogeneous due to the micromagnetic configuration. The magnitude of the $\bm{h}$ field can also vary due to a nonuniform coupling strength. Moreover, recent theoretical investigations of ferromagnetic InAs-Al-EuS nanowires showed that the electrostatic environment is crucial in modulating the effect of the EuS on the InAs \cite{Escribano_PRB_2021, Liu_PRB_2021, Escribano_NPJQM_2022}, suggesting that, in principle, it is possible to tune the induced exchange field electrostatically.

In this work, we consider that the exchange field $\bm{h}$ takes a constant value $\bm{h}_j$ in each of the three regions $i\in \{\mathrm{L}, \mathrm{C}, \mathrm{R}\}$. We will call collectively $\mathrm{L}$ and $\mathrm{R}$ lateral regions, with an exchange field $\bm{h}_\mathrm{L}=\bm{h}_\mathrm{R}=\bm{h}_\mathrm{lat}$ while we will use the symbol $\bm{h}_\mathrm{all}$ when considering a homogeneous value for the exchange field.
We use the same notation for the chemical potential $\mu$ of the three regions. In addition, we introduce potential barriers with height $V_\mathrm{B}$ at the interfaces between the central and the lateral regions, tuning the system from the open ($V_\mathrm{B} = 0$) to the quantum dot regimes ($V_\mathrm{B} \gg -\mu_\mathrm{lat}$). The induced pairing potential is $\Delta_j = \Delta_{0, j} e^{i \phi_j} $ with the modulus $\Delta_{0, j}$ taking finite value only in the $\mathrm{L}$ and $\mathrm{R}$ regions while the superconducting phase difference between the two leads is $\phi = \phi_\mathrm{R} - \phi_\mathrm{L}$.

In all the simulations we use realistic parameters for InAs-Al-EuS heterostructures, taking $\Delta_0 = \SI{0.250}{\milli\eV}$ and effective mass $m^* = 0.026\,m_e$. We consider a nanowire of total length including both the lateral and central regions of $\ell_\mathrm{W} = 3~\mu$m. To obtain numerical results, we discretize the Hamiltonian using a finite-differences scheme with a lattice spacing of $a = \SI{2}{\nano\meter}$ implemented using the \texttt{KWANT} package \cite{Groth_NJP_2014}. The code and the results of the simulations are available at~\cite{repo}.

We focus on the short junction limit, and we consider a central region of length  $\ell_\mathrm{C}=\SI{180}{\nano\meter}$, 
such that the quantization energy is comparable to the other energy scales. We fix the chemical potential in the lateral regions to be $\mu_\mathrm{L} = \mu_\mathrm{R} = 16~\Delta = \SI{4}{\milli\eV}$. Directly solving for the quasiparticle spectrum of the continuum model allows treating on equal footing the current carried by Andreev bound states (ABSs) and the quasi-continuum of states above the gap. Usually, the continuum current is subdominant, except for strong exchange fields, where its contribution is comparable to or even larger than the ABS one~\cite{Krichevsky_PRB_2000, Levchenko_PRB_2006, Benjamin_EPJB_2007, Bujnowski_EPL_2016, Razmadze_2023}.

The supercurrent in a Josephson junction is an equilibrium phenomenon that can be described by a function $E_\mathrm{J}(\phi)$, called Josephson potential or phase dispersion relation. This can be evaluated from the quasiparticle spectrum, assuming that the quasiparticles are in thermal equilibrium and calculating the free energy at a fixed phase 
\begin{equation}
\begin{split}
    E_\mathrm{J}(\phi) &= - k_\mathrm{B} T_\mathrm{p} \ln \tr e^{-\frac{H}{k_\mathrm{B} T_\mathrm{p}}} \\ &= - k_\mathrm{B} T_\mathrm{p} \sum_n \ln \Big[2 \cosh (\frac{\omega_n(\phi)}{2 k_\mathrm{B} T_\mathrm{p}})\Big]\,,
\end{split}
    \label{eq:free_energy}
\end{equation}
where  $\omega_n$ is the quasiparticle spectrum, $k_\mathrm{B}$ is the Boltzmann constant and $T_\mathrm{p}$ is the quasiparticle  temperature~\cite{Kosztin_PRB_1998}. From $E_\mathrm{J}$, the CPR is calculated through the thermodynamic relation 
\begin{equation}
    \langle I \rangle = \frac{2 e}{\hbar} \pdv{E_\mathrm{J}(\phi)}{\phi}\,.
\label{eq:current}
\end{equation}

The maximum current that can flow in the junction in equilibrium is called critical current, $I_\mathrm{c}$, while we define the critical phase, $\phi_\mathrm{c}$, as the phase difference where this is reached 
\begin{equation}
    I_\mathrm{c} = |I (\phi_\mathrm{c})|,\qquad \phi_\mathrm{c} \equiv \argmax_{\phi\in[0, \pi]} \abs{I(\phi)}\,,
\end{equation}
where we restricted the search domain to $[0, \pi]$ in the reciprocal case. We define the quantity $I_0 \equiv2  e \Delta/\hbar$ as the relevant current scale that has the numerical value $I_0 = \SI{122}{\nano\ampere}$ for Al. 
The sign of $I(\phi_\mathrm{c})$ defines the direction of the supercurrent at the critical phase. 

It is useful to decompose the phase dispersion in its harmonic components
\begin{equation}
    E_\mathrm{J}(\phi) = \sum_{k=1}^{\infty} [C_k \cos(k \phi) + S_k \sin(k \phi)]\,,
    \label{eq:harmonic_decomposition}
\end{equation}
since each harmonic corresponds to the tunneling of multiplets of Cooper pairs between the two superconducting regions. To see this, it is necessary to interpret the phase in Eq.~\eqref{eq:harmonic_decomposition} as an operator and rewrite the expression in the charge basis by interpreting $\exp(i k \phi)$ as a translation operator. Neglecting the constant term, the result is
\begin{equation}
    \mathcal{H}_{J} = \sum_n \sum_k \frac{C_k + i S_k}{2} \ket{n+k}\bra{n} + \mathrm{H.c.}
\end{equation}
where $\ket{n}$ is the state with a difference of $n$ Cooper pairs in the two leads. In this way, it is easy to see how the terms $\cos(k \phi)$ mediate the tunneling of $k$ Cooper pairs. 

A purely $\cos(2\phi)$ Josephson junction can be used to create a qubit with nearly-degenerate ground states that are a superposition of only states with the same parity of Cooper pair number.~\cite{Smith_npjQI_2020, Guo_PRB_2022}. In this setup, $\cos(\phi)$ and $\sin(\phi)$ perturbations can be used to implement rotation in the Bloch sphere~\cite{Maiani_PRXQ_2022}.

If time-reversal and inversion symmetries are not simultaneously broken, the Josephson junction is reciprocal and $E_\mathrm{J}(\phi) = E_\mathrm{J}(-\phi)$. It subsequently results in the absence of the ${S_k}$ components, called anomalous. These anomalous components are necessary for  $\phi_0$ junctions and the diode effects. 

In the reciprocal case, assuming all the components $\{C_k\}$ with $k>2$ are negligible, the only minima of the Josephson potential are located at $\phi = 0$ and $\pi$ only if $|C_{1}| \geq 4 C_{2}$.  For $|C_{1}| < 4 C_{2}$, minima can be found at $\phi = \pm \arctan (C_1/\sqrt{16 C_2^2-C_1^2})$. This case, dubbed $\pm \phi_0$-junction, does not require an inversion-symmetry breaking mechanism~\cite{Goldobin_PRB_2007}.

{ Throughout this study, the dominant harmonic is consistently either the fundamental or the second one, with subdominant contributions from third and higher harmonics. Nonetheless, all calculations incorporate their effects, even though they are not showcased in any graphs.}

\subsection{Single level model}
\label{sec:single_level}

\begin{figure}[t!]
    \centering
    \includegraphics[width=0.95\columnwidth]{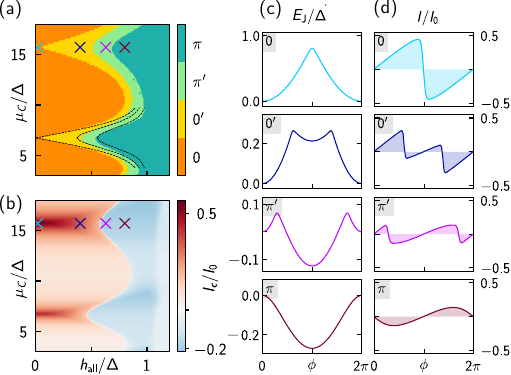}
    \caption{
    \textbf{Phase diagram in the quantum dot regime.}
    (a) Phase diagram, and (b) critical current for a system in the dot regime as a function of the chemical potential in the central region, $\mu_\mathrm{C}$, and the exchange field, $h_\mathrm{all}$, considered homogeneous in the heterostructure. The dashed lines are an overlay of the analytical model in Eq.~\eqref{eq:zero_crossings} where we selected $\varepsilon=16~\Delta$ and $\gamma=0.2~\Delta$. (c) and (d) show the Josephson potential and the CPR for some selected points in the parameter space [crosses in (a) and (b)].
    Parameters for the continuum BdG model: $\mu_\mathrm{lat} = \SI{4}{\milli\eV}$, $\ell_\mathrm{B}=\SI{15}{nm}$, $V_\mathrm{B} = \SI{3}{\milli\eV}$.
    }
\label{fig:dot-phases}
\end{figure}

Before proceeding to the numerical results, we introduce a single-level model in which a level with energy $\varepsilon$ couples to two high carrier density superconducting lateral regions, see App.~\ref{sec:app:gfmodel} and Ref.~\cite{Razmadze_2023}.

To get an expression for the critical lines separating the different phases, we consider the condition where an ABS crosses the Fermi level, obtained from Eq.~\ref{eq:app:abs} for $\omega=0$. When a spin-split ABS crosses the Fermi level, its occupation changes and provides no contribution to the current. Therefore, the residual current is entirely due to the continuum of states and has a characteristic $\pi$ contribution~\cite{Razmadze_2023}.

In the case of equal exchange fields in the leads, $h_\mathrm{L} = h_\mathrm{R} = h_\mathrm{lat}$, the Fermi level crossing condition for ABS with spin $\sigma=\pm1$ is given by the expression
\begin{equation}
\begin{split}
\frac{\sigma h_\mathrm{C}}{\gamma} + &\frac{\sigma h_\mathrm{lat}}{\sqrt{\Delta^2-h_\mathrm{lat}^2}} = \\
\pm &\sqrt{\frac{\varepsilon^2}{\gamma^2} + \frac{\Delta^2}{\Delta^2-h_\mathrm{lat}^2}\Big[ 1 - T \sin^2(\phi/2)\Big]}\,,
\end{split}
\label{eq:zero_crossings}
\end{equation}
where $\gamma=\gamma_\mathrm{L}+\gamma_\mathrm{R}$ is the tunnel rate to the leads and $T= 4 \gamma_\mathrm{L} \gamma_\mathrm{R}/(\gamma_\mathrm{L}+\gamma_\mathrm{R})^2$ is the transparency of the junction. When a spin-split ABS crosses the Fermi level at $\phi=\pi$, a metastable $\pi$ phase appears, marking the transition from $0$ to $0'$. When such a crossing happens for $\phi=0$, the $0$ phase becomes completely unstable, marking the $\pi'$ to $\pi$ transition. Finally, the $0'$ to $\pi'$ critical line can be approximated by $\phi=\pi/2$.

The quantum point contact limit, in which the intermediate state in the junction is strongly hybridized with the states in the leads, can be obtained from Eq.~\ref{eq:app:abs} for $\gamma\to\infty$. This results in a generalization of Beenakker's formula~\cite{Beenakker_PRL_1991b} for spin-split leads
\begin{equation}
\omega = \pm \Delta \sqrt{1-T \sin^2(\phi/2)} - \sigma h_\mathrm{lat}\,.
\end{equation}

\section{Controllable CPR}
\subsection{Quantum dot regime}
\label{sec:quantum_dot}
The quantum dot regime is reached when large barriers at the edge of the central region are introduced, see Fig.~\ref{fig:sketch}. In this regime, electrons are confined in the central region. The quantum dot regime is optimal for the electrostatic controllability of the 0\,--\,$\pi$ transition. When the quantum dot levels align with the chemical potential in the leads, the $\pi$ phase appears at low exchange fields, and the critical line takes the form $h_\mathrm{C} \propto \varepsilon$, see Fig.~\ref{fig:dot-phases}(c). This work does not consider the electrostatic repulsion in the central region (Coulomb blockade). We expect the mean-field picture to be valid for large exchange fields, and the Coulomb repulsion to enhance the exchange field in the central region.

In this regime, $I_\mathrm{c}$ is maximal when the dot levels align with the chemical potential of the leads,  as shown in Fig.~\ref{fig:dot-phases}(b). In the off-resonance condition, the critical lines converge to the spectral gap closing point, $h=\Delta$. The phase diagram can be understood using the single-level model, which predicts the hyperbolic critical lines [see the dashed lines in Fig.~\ref{fig:dot-phases}(a)]. The Josephson potential and CPRs on resonance are shown in Figs.~\ref{fig:dot-phases}(c) and (d). The fundamental harmonic dominates the junction properties in both the 0 and the $\pi$ phases (top and bottom panels). In contrast, high-harmonic contributions become important in the $0'$ and $\pi'$ phases because of the double minima Josephson potential.

\begin{figure}[t!]
    \centering
    \includegraphics[width=\columnwidth]{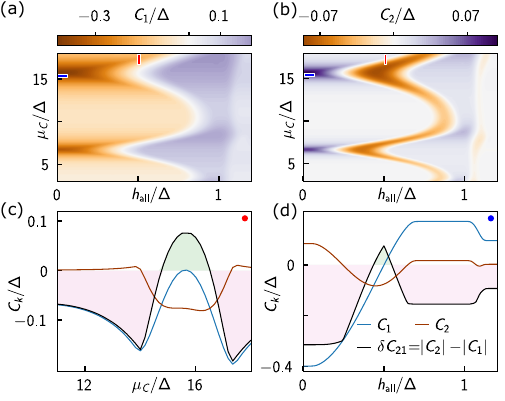}\caption{
    \textbf{Harmonic components in the dot regime.}
    Fundamental (a) and second (b) harmonic for a ferromagnetic hybrid junction in the dot regime.
    The transition from 0 to $\pi$ appears close to the gap closing ($h/\Delta=1$) for a detuned dot, while, near resonance, the $\pi$ phase can appear at lower magnetic fields. In the regions of the metastable phases, a strong $C_2$ component can be observed while the $C_1$ component vanishes. In panels (c) and (d), we show two cuts of the harmonic components [blue and red ticks in (a) and (b)]. For constant $h_\mathrm{all}$ the $C_k$ components show a peak when a dot level crosses the Fermi level of the leads, but the width is increasingly smaller for higher harmonics. For constant $\mu_\mathrm{C}$, the CPR show a sinusoidal behavior until the system reaches the 0--$\pi$ transition, where the second harmonic dominates. The parameters are equal to those in Fig.~\ref{fig:dot-phases}.
    }
    \label{fig:dot}
\end{figure}

To better understand the harmonic composition of the Josephson energy, we present the lowest harmonic components of the CPR in Fig.~\ref{fig:dot}. For a small value of the exchange field, in the $0$ and $0'$ phases, the harmonic component coefficients show a peak corresponding to an energy level in the quantum dot aligning with the Fermi level of the lateral regions. These peaks have widths that decrease for higher-order components, allowing the relative strength of the first two harmonics, $\delta C_{21} = |C_2| - |C_1|$, to be tuned by slightly changing the energy of the quantum dot levels electrostatically. In contrast, the sensitivity of $E_\mathrm{J}$ to the chemical is almost negligible in the $\pi$ phase and the suppression of $C_2$ is significant, resulting in a sinusoidal CPR, as shown in Fig.~\ref{fig:dot-phases}(d). When the system is tuned to the vicinity of the $0'$--$\pi'$ transition, the fundamental harmonic is suppressed, leading to a regime dominated by the second harmonic and a double-well Josephson potential.

\subsection{Open regime}
\begin{figure}[t!]
    \centering
    \includegraphics[width=\columnwidth]{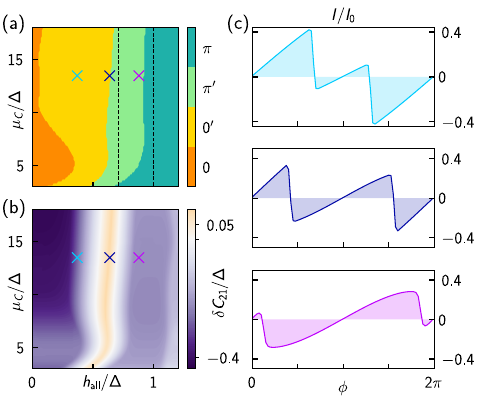}\caption{\textbf{Open regime.}
    (a) Phase diagram and (b) difference between the second and first CPR harmonic component for a ferromagnetic hybrid junction in the open regime. The overlaid dashed lines are the prediction of the analytic model in the $\gamma \to \infty$ limit and $T=1$. The sensitivity of the phase boundaries to $\mu_\mathrm{C}$ is strongly suppressed in this limit, while the phase space occupied by $0'$ and $\pi'$ phases is increased. Parameters: $\mu_\mathrm{lat} = \SI{4}{\milli\eV}$, $\ell_\mathrm{B} = \SI{0}{\nano\meter}$.
    \label{fig:open}
    }
\end{figure}

In contrast to the quantum dot regime, the open regime shows a weak dependence on chemical potential, except close to the edge of the band ($\mu=0$), Fig.~\ref{fig:open} (a). The open regime shows a $\pi$ phase for $h_{\rm all}>\Delta$, as predicted by the analytic expression in Eq.~\eqref{eq:zero_crossings}. The system shows extended metastable $0'$ and $\pi'$ regions compared to the quantum dot regime. In the open regime, the transition happens for $h_\mathrm{all}=\Delta/\sqrt{1-T/2}$. For transparent junctions, this critical line coincides with the zero-temperature paramagnetic limit for superconductors meaning that this regime cannot be achieved in materials with intrinsic superconductivity. Instead, semiconductor-superconductor devices are ideal for reaching the $\pi'$ and $\pi$ regimes. At the $0'$--$\pi'$ transition point, the CPR is dominated by the $\sin(2 \phi)$ term, leading to a Josephson potential with two equivalent minima within the $\phi\in[0,\pi]$ range. We note that the robustness against local fluctuations in $\mu_\mathrm{C}$ is a unique feature of the open regime. The other two transitions lines, for $0$--$0'$ and $\pi$--$\pi'$, are also almost independent of the chemical potential once $\mu_C\gtrsim10$, taking place close to $h_\mathrm{all}=0$ and $h_\mathrm{all}=\Delta\sqrt{1-T}$. 

In both open and dot cases, when $C_1$ reaches the crossover point and becomes zero, $C_2$ has a negative sign. This satisfies the condition $|C_1| > 4C_2$ discussed before, resulting in minima at $0$ and $\pi$. For what concerns  temperature dependence, since the higher harmonic components are dictated mainly by the lowest ABS, the main effect of increasing temperatures is a reduction of the metastable $0'$ and $\pi'$ regions.

The open and dot regimes have different advantages and disadvantages for practical applications of ferromagnetic hybrid junctions as a $\cos(2\phi)$ Josephson element. The open regime is insensitive to noise in the gate voltage but requires a relatively high exchange field $h_\mathrm{all} \simeq \Delta / \sqrt{2}$ for the second harmonic to dominate. Increasing the barriers, and thus moving toward the dot regime, lowers the required exchange field toward the theoretical limit $h_\mathrm{all}\simeq0$ at the price of a higher sensitivity to gate noise. It also allows electrostatic control of the harmonic content. Indeed, when the $C_2$ component approaches its maximum location, $C_1$ vanishes linearly. This opens up the possibility of introducing a gate-controllable $\cos(\phi)$ component. Additionally, the ability to change from a dot to an open regime is controlled by the electrostatic environment, allowing to tune the system between the two regimes.

\subsection{Inhomogeneous exchange field}
\begin{figure}[t!]
    \centering
    \includegraphics[width=\columnwidth]{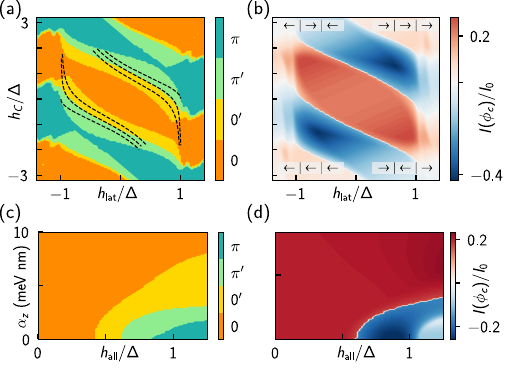}
    \caption{
    \textbf{Effect of inhomogeneous exchange field and spin-orbit coupling.} 
    Phase diagram (a) and critical current (b) for a ferromagnetic hybrid junction with an inhomogeneous exchange field. The overlaid dashed lines are the prediction of the single-level model, Eq.~\eqref{eq:zero_crossings}. The system shows the alternation of $0$ and $\pi$ phases as a function of $h_\mathrm{C}$ with sharp transitions.  A transition from $0$ to $\pi$ can also be obtained by gap closing ($h / \Delta = 1$). 
    Phase diagrams (c) and critical current (d) of the ferromagnetic hybrid junction in the presence of spin-orbit coupling such that $\bm{\kappa}\cdot\bm{h}=0$. Spin-orbit coupling leads to general suppression of the $\pi$ phase and the expansion of the metastable phases.
    Parameters: $\mu_\mathrm{C}=\SI{1}{\milli\eV}$, $\mu_\mathrm{lat}=\SI{4}{\milli\eV}$, $\ell_\mathrm{B} = \SI{0}{\nano\meter}$.
    }
    \label{fig:inhomogenous_h_and_soc}
\end{figure}

The exchange field in the lateral and central regions affects CPR differently. To reveal this difference, we now analyze the case  where the exchange field in the central ($h_\mathrm{C}$) and lateral regions ($h_\mathrm{R}=h_\mathrm{L}=h_\mathrm{lat}$) have different values while being still aligned in the same direction, see Fig.~\ref{fig:inhomogenous_h_and_soc}(a) and (b). In the case of small magnetization in the lateral regions ($h_\mathrm{lat}\sim0$), we find that a strong polarization in the central one $h_\mathrm{C}\gg\Delta$ is needed to induce the transition to the $\pi$ state in the open regime. 

The exchange field strength necessary to induce a $0$\,--\,$\pi$ transition crucially depends on other parameters of the system. In particular, longer junctions and low density in the central region are associated with transitions at lower fields. The length dependence can be understood using a semiclassical model, where $h_\mathrm{C}$ adds an extra phase accumulated by quasiparticles in a round-trip between the leads. This phase is proportional to $h_\mathrm{C} \,\ell_\mathrm{C}$ product, explaining why longer junctions exhibit switches from the 0 to the $\pi$ phase at lower exchange field values. In addition, it leads to a periodic pattern of $0$ and $\pi$ phases along the $h_\mathrm{C}$ axis. A similar effect can be obtained by reducing $\mu_\mathrm{C}$, which reduces the Fermi velocity. This is further discussed in App.~\ref{sec:app:semiclassical}.

A sharp transition from $0$ to $\pi$ can also be obtained near gap closing ($\abs{h_\mathrm{lat}}=\Delta)$. In this case, the transition is associated with a strong reduction in the magnitude of $I_\mathrm{c}$. This behavior can be understood using the simplified one-level model in Eq.~\eqref{eq:zero_crossings}, which explains the transition to the $\pi$ phase as the disappearance of the contribution of the lowest ABS, thus reducing the total supercurrent in the junction. The fact that the 0\,--\,$\pi$ transition is associated with a decrease of critical current only in the case of gap closing can be potentially used to infer the dominant mechanism in experiments.

\subsection{Spin-orbit coupling in the semiconductor}
\label{sec:soc}
So far, we have neglected the effect of spin-obit coupling. Here, we consider a simple model for linear spin-orbit coupling that, for a quasi-1D system, takes the form 
\begin{equation}
\mathcal{H}_\mathrm{SOC} =  k_x [\alpha_z \sigma_y + \beta \sigma_x] \tau_z = k_x [\bm{\kappa} \cdot \bm{\sigma}] \tau_z \,,
\label{eq:soc_hamiltonian}
\end{equation}
where we defined a spin-orbit coupling vector $\bm{\kappa}= (\beta_x, \alpha_z, 0)$~\cite{Maiani_2022}. In the simplest setup, $\alpha_z$ arises from the Rashba field and $\beta$ from the Dresselhaus term. We note that the distinction between the two terms is artificial in a one-dimensional model, as the two terms can be mapped onto each other by a unitary transformation
\begin{equation}
    U = \exp( -i \theta/2 \sigma_z \tau_0)\,,
    \end{equation}
that is a rotation in spin space around the $z$ axis by an angle $\theta$, potentially inhomogeneous in space. Using $\theta=\arctan(\beta/\alpha_z)$, we can always remove the term proportional to $\sigma_x$ and align the spin-orbit vector in the $y$ direction. This transformation, however, also rotates the exchange field. This result illustrates the equivalence between inhomogeneous spin-orbit fields and exchange fields.

We first focus on the homogeneous spin-orbit situation, displayed in Fig.~\ref{fig:inhomogenous_h_and_soc}~(c) and (d). Although spin-orbit coupling splits the Fermi surface, Cooper pairs do not acquire a finite momentum unless time-reversal symmetry is broken. Thus, the oscillation between triplet and singlet components is absent, and it is impossible to obtain a $\pi$ phase. At a finite magnetic field,  the Rashba term couples the two spin-split ABSs reopening the gap, unless the field aligns with the spin-orbit vector $\bm{\kappa}$. The effect on the CPR of a transverse Rashba field is a substantial reduction of the $\pi$ regions and an enlargement of the metastable phases. When the exchange field $\bm{h}$ is instead aligned with $\bm{\kappa}$, the spin-rotation symmetry is unbroken. This allows for $\pi$ phases, but simultaneously, the system remains gapless for $h>\Delta$. 

Anomalous Josephson effects can occur when a spin-orbit coupling vector is aligned with a magnetic field~\cite{Buzdin_PRL_2008}. However, it is not observed in the homogeneous case as the combination of various spin-rotation symmetry-breaking effects is necessary for its manifestation~\cite{Rasmussen_PRB_2016}.
This can occur, for instance, due to finite spin-splitting with a non-zero component in both the junction direction and the transverse one~\cite{Baumard_PRB_2020}, or multiple modes that can hybridize~\cite{Zazunov_PRL2009}. 
The presence of anomalous currents in similar systems has been considered in Refs.~\cite{Cheng_PRB_2012, Nesterov_PRB_2016}.

\begin{figure}[t!]
    \centering
    \includegraphics[width=\columnwidth]{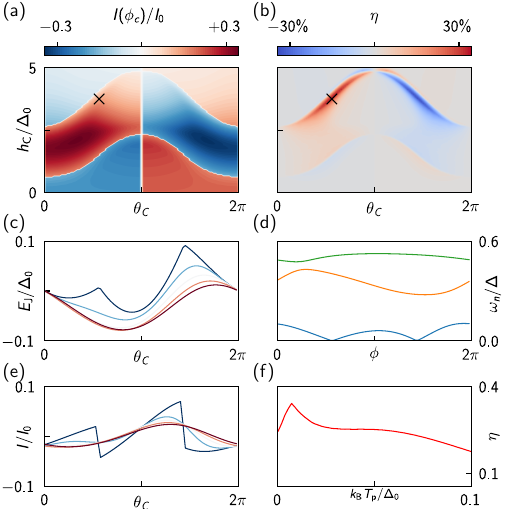}
    \caption{\textbf{Non-reciprocal behavior.}
    We introduce a spin-orbit coupling $\kappa =\SI{2}{\nm \eV}$, misaligned with an angle $\theta_\mathrm{C}$ between the central and the lateral regions. Panel (a) shows the current at the critical phase, while the diode efficiency is shown in (b). (c) The Josephson potential and (e) CPR for a specific point of the parameter space [black cross in (a)] are displayed for increasing temperature from zero to \SI{0.5}{K}. Non-reciprocal behavior shows a non-monotonic temperature dependence, evident in panel (f). This can be explained by the ABS spectrum [panel (d)] that comprises a reciprocal lowest state. Therefore, increasing the temperature suppresses the reciprocal contribution increasing the efficiency.
    Parameters: $\mu_\mathrm{C}=\SI{1}{\milli\eV}$, $\mu_\mathrm{lat}=\SI{4}{\milli\eV}$, $\ell_\mathrm{B} = \SI{0}{\nano\meter}$,  $h_\mathrm{lat} = 0.8\Delta$, $\kappa = \SI{2}{\milli\eV \nano\meter}$.
    }
    \label{fig:aje}
\end{figure}

The spin-orbit field depends on the local electrostatic environment. For this reason, the spin-orbit direction can have different magnitudes and directions in different regions of the ferromagnetic hybrid junction. We consider this situation in Fig.~\ref{fig:aje}.
The spin-orbit direction is misaligned in the central region by an angle $\theta_\mathrm{C}$ with respect to the lateral ones, while we consider a homogeneous exchange field. Since the Rashba field is proportional to the electric field, this scenario might appear in Josephson junctions due to a varying electrostatic environment. This is equivalent to a homogeneous spin-orbit coupling field and a misaligned exchange field in the three regions. In this case, the CPR shows a non-reciprocal behavior, $I(\phi)\neq I(-\phi)$, due to the presence of anomalous $\sin(k \phi)$ terms in the Josephson potential. The non-reciprocal supercurrent has been recently reported in superconductor-semiconductor nanowires~\cite{Mazur_arXiv2022}.
The critical phase in the non-reciprocal case reads as $\phi_\mathrm{c} \equiv \argmax_{[0, 2\pi)} \abs{I}$. To measure the non-reciprocal behavior, we define the critical current in the two directions, $I_\mathrm{c}^+\equiv\max I$ and  $I_\mathrm{c}^-\equiv-\min I$, and the corresponding diode efficiency as $\eta \equiv (I_\mathrm{c}^+ - I_\mathrm{c}^-)/(I_\mathrm{c}^+ + I_\mathrm{c}^-)$. For the parameters considered, the efficiency can be as high as $30\%$ in the region close to the 0\,--\,$\pi$ transition. For this point, the CPR shows a characteristic form $I(\phi) \sim \sin(\phi - \phi_0) + \cos(2\phi)$. 

The rectification effect exhibits a non-monotonic temperature dependence $\eta(T_p)$ that shows a maximum at finite temperature. This can be explained by the ABS spectrum, see Fig.~\ref{fig:aje} (d): the lowest state is dominated by the $\cos(k\phi)$ contribution, with a weak non-reciprocal behavior. The first excited state has a predominant $\sin(\phi)$ contribution. Therefore, increasing the temperature increases the non-reciprocal supercurrent contribution through an interference process of the different CPR harmonics~\cite{Souto_PRL_2022}. At higher temperatures, more states become populated, washing out the contribution from the harmonics and leading to a sinusoidal CPR, Fig.~\ref{fig:aje}(e).

\begin{figure}[t]
    \centering
    \includegraphics[]{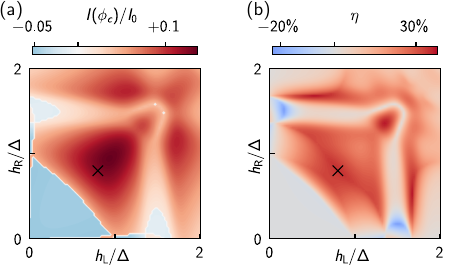}
    \caption{\textbf{Non-reciprocal behavior for junctions with asymmetric exchange fields.}
    Panel (a) shows the supercurrent at the critical phase, while the diode efficiency is shown in (b). The black cross represents the point of maximum efficiency of Fig.~\ref{fig:aje}.
    Parameters: $\mu_\mathrm{C}=\SI{1}{\milli\eV}$, $\mu_\mathrm{lat}=\SI{4}{\milli\eV}$, $\ell_\mathrm{B} = \SI{0}{\nano\meter}$, $\kappa = \SI{2}{\milli\eV \nano\meter}$, $\theta_\mathrm{C} = 0.56 \pi$ .
    }
    \label{fig:asymmetric_diode}
\end{figure}

{
The nonreciprocal behavior persists for Josephson junctions that feature a different magnitude of the exchange field in both leads, with a diode efficiency that even increases in some cases. This is illustrated in Fig.~\ref{fig:asymmetric_diode}, where we take the parameters of the maximum efficiency in Fig.~\ref{fig:aje}, marked by the black cross, and tune the magnitude of the exchange fields in both leads, $h_\mathrm{L}$ and $h_\mathrm{R}$, independently.
}

\section{Conclusions}
In this work, we have analyzed the higher harmonics in the current-phase relation appearing in ferromagnetic hybrid junctions. The induced exchange field in the semiconductor can overcome the induced pairing potential without causing a transition to the normal state. This results in spin-polarized Andreev bound states with opposite spin crossing the Fermi level, leading to the 0\,--\,$\pi$ phase transition and supercurrent reversal. 

Higher harmonics in the current-phase relation become dominant close to the 0\,--\,$\pi$ transition, where the supercurrent changes sign. In the dot regime (weak coupling to the leads), the $\pi$ depends on the relative position of the dot levels with respect to the leads chemical potential. In case the junction is tuned into the open regime (large coupling to the leads), the onset of the $\pi$ phase is less sensitive to changes in chemical potential. 

We find that the spin-orbit coupling increases the coexisting region between $0$ and $\pi$ phases with considerable amplitudes of higher-harmonic components. Finally, we find that non-collinear spin-orbit coupling in the junction, due to a varying electrostatic environment, results in a supercurrent rectification effect whose efficiency peaks around 0\,--\,$\pi$ transition.

{
Despite the model's simplicity, it captures the main aspects of physical phenomenology
This becomes evident when analyzing extreme scenarios. If the two barriers are finite, a quantum dot is formed, resulting in a low-field $0$\,--\,$\pi$ transition that agrees well with previous experiments~\cite{Razmadze_PRL2020, VanDam_Nature_2006}. In the case of single-level regimes, a significant spin splitting from either magnetic proximity or field is required to induce the $\pi$ phase, consistent with previous findings~\cite{Razmadze_2023, Whiticar_2021, Ke_NatCom_2019}. 

For ferromagnetic hybrid junctions, the CPR is determined by the domain configuration. Therefore, achieving the $0'$ and $\pi'$ phases would require control of the magnetic properties of the materials involved. Deterministic individual domain flipping might be complicated to control, but there is evidence~\cite{Razmadze_2023} that it occurs naturally in nanowires.  

Gate-tunability of the charge density in a ferromagnetic hybrid platform is readily available. Electrostatic control of the ferromagnetic proximity effect, which has been already demonstrated in superconductor-ferromagnetic insulator heterostructures~\cite{Liu_2013}, would greatly increase the attractiveness of the platform by making it possible to have significant control over the CPR.
}

The tunability of the harmonic content of CPRs is relevant for a number of applications, including superconducting diodes~\cite{Silaev_JPCM2014, Yokoyama_PRB2014, Halterman_PRB2022, Ilic_PRL2022, Ilic_PRAp2022, Davydova_SciAdv2022, Souto_PRL_2022, Mazur_arXiv2022,soori2023anomalous,steiner2022diode,tanaka2022theory,kokkeler2022field,haenel2022superconducting,fu2022gate, Trahms_2023,turini2022josephson,daido2022superconducting,lu2022josephson,wei2022supercurrent,fominov2022asymmetric,chiles2022non,legg2022superconducting,wang2022symmetry,song2022interference,gupta2022superconducting, Soori_2023, Ortega_Taberner_2023, Hu_diodeArxiv,legg2023parity}, ferromagnetic transmon qubits~\cite{Ahmad_PRB_2022}, and parity-protected qubits~\cite{Smith_npjQI_2020, Larsen_PRL_2020, Guo_PRB_2022, Schrade_PRXQ_2022, Maiani_PRXQ_2022}. In this context, ferromagnetic junctions in the open regime are promising thanks to their robustness against both charge and flux fluctuations.

\section{Acknowledgments}
This work was supported by the Danish National Research Foundation, the Danish Council for Independent Research \textbar Natural Sciences, Swedish Research Council (VR), the European Research Council (ERC) under the European Union’s Horizon 2020 research and innovation program under Grant Agreement No. 856526, and Nanolund. We also acknowledge support from Microsoft and a research grant (Project 43951) from VILLUM FONDEN.

\appendix
\section{Analytical model for the short-junction limit}
\label{sec:app:gfmodel}
In this appendix, we introduce a simple analytical model that describes the transport through the system, employing the Green function formalism outlined in Ref.~\cite{Rodero_AP_2011}.

The retarded/advanced (R/A) Green function of the central region is given by
\begin{equation}
    \hat{G}^{R/A}_0(\omega)=\left[\hat{g}_{0}^{-1}(\omega)-\hat{\Sigma}^{R/A}_\mathrm{L}(\omega)-\hat{\Sigma}^{R/A}_\mathrm{R}(\omega)\right]^{-1},
    \label{Eq:Dyson}
\end{equation}
where $\hat{g}_{0}^{-1}(\omega)=(\omega+\sigma h_{\rm C})\hat{\tau}_0+\varepsilon\hat{\tau}_z$ describes the isolated normal region as a function of the electron energy $\omega$, and $\hat{\Sigma}^{R/A}_\nu$ is the self-energy describing the coupling to the lateral regions $\nu$.

In the wide bandwidth limit, the self-energy of these regions is given by
\begin{equation}
    \hat{\Sigma}^{R/A}_{\nu}(\omega)=\sum_\nu\Gamma_\nu\left[\mathfrak{g}^{R/A}_{\nu}(\omega)\hat{\tau}_0+\mathfrak{f}^{R/A}_{\nu}(\omega)\tau_x e^{\hat{\tau}_y\phi_\nu}\right]\,.
\end{equation}
Here 
\begin{align}
    \mathfrak{g}^{R/A}_{\nu}(\omega)&=-\frac{\omega+\sigma h_{\nu}\pm i\eta}{\sqrt{\Delta^2-(\omega+\sigma h_{\nu}\pm i\eta)^2}}\,,\qquad \\
    \mathfrak{f}^{R/A}_{\nu}(\omega)&=\frac{\Delta}{\sqrt{\Delta^2-(\omega+\sigma h_{\nu}\pm i\eta)^2}}\,,
    \label{Eq::GFs}
\end{align}
while $\eta$ is the Dynes parameter, controlling the width of the superconducting coherent peaks at $\omega=\pm\Delta$, which we take infinitesimal in this case.

\begin{widetext}
In the short junction case, we can use a minimal model where the central region is described by a single electronic site with energy $\varepsilon$ and exchange splitting $h_\mathrm{C}$. To determine the ABS spectrum, we search for the poles of the retarded Green's function by solving $\det[(G^R)^{-1}]=0$. This results in the complicated expression
\begin{equation}
\Bigg(\omega + \sigma h_\mathrm{C} + \sum_\nu \gamma_\nu \frac{\omega + \sigma h_\nu}{\sqrt{\Delta^2-(\omega+\sigma h_\nu)^2}}\Bigg)^2 = \varepsilon^2 + \abs{\sum_\nu \frac{\Delta \gamma_\nu e^{i\phi_\nu}}{\sqrt{\Delta^2-(\omega+\sigma h_\nu)^2}}}^2    
\end{equation}

The expression becomes more compact if we consider $h_\mathrm{R}=h_\mathrm{L}=h_\mathrm{lat}$. In this case, the expression simplifies to 
\begin{equation}
\Bigg(\omega + \sigma h_\mathrm{C} + \sum_\nu \gamma_\nu \frac{\omega + \sigma h_\mathrm{lat}}{\sqrt{\Delta^2-(\omega+\sigma h_\mathrm{lat})^2}}\Bigg)^2 = \varepsilon^2 + \frac{\Delta^2}{\Delta^2-(\omega+\sigma h_\mathrm{lat})^2} b^2(\phi)    
\end{equation}
where we defined the phase potential
\begin{equation}
    b(\phi) = \abs{\sum_\nu \gamma_\nu e^{i\phi_\nu}}\,.
\end{equation}
Note that with this minimal simplification, the dependence on the phase is entirely condensed in the function $b(\phi)$.

By defining the total coupling $\gamma=\gamma_\mathrm{L} + \gamma_\mathrm{R}$ and the transmission as $T = 4 \gamma_\mathrm{L} \gamma_\mathrm{R} / (\gamma_\mathrm{L} + \gamma_\mathrm{R})^2$, we get a simplified expression that reads as
\begin{equation}
\frac{\omega + \sigma h_\mathrm{C}}{\gamma} + \frac{\omega + \sigma h_\mathrm{lat}}{\sqrt{\Delta^2-(\omega+\sigma h_\mathrm{lat})^2}} = \pm \sqrt{\frac{\varepsilon^2}{\gamma^2} + \frac{\Delta^2}{\Delta^2-(\omega+\sigma h_\mathrm{lat})^2}\Big[ 1 - T \sin^2(\phi/2)\Big]}\,. 
\label{eq:app:abs}
\end{equation}
\end{widetext}

The quantum point contact limit is defined for $\gamma\to\infty$ and results in a generalization of Beenakker's formula for spin-split leads
\begin{equation}
\omega = \pm \Delta \sqrt{1-T \sin^2(\phi/2)} - \sigma h_\mathrm{lat}\,,
\end{equation}
while the case for $\omega=0$ gives the Fermi level crossing formula used in Eq.~\eqref{eq:zero_crossings}.

\section{Semiclassical analysis of a long junction}
\label{sec:app:semiclassical}
The more conventional regime is described by the energy scale hierarchy $\mu \gtrsim \Delta \gtrsim h$ in the lateral regions, and $\mu \gtrsim h$ in the central one. In this case, the device behaves substantially as a superconductor-ferromagnet-superconductor junction. For long junctions with a sufficiently high density and small polarization, the behavior can be approximately described by a semiclassical model. This assumption is not met in the system studied by the numerical model. Nevertheless, this approach provides a clearer qualitative picture of the physics of this system. We follow the steps of Refs. \cite{Beenakker_PRL_1991b, Bagwell_PRB_1992} including a spin splitting oriented in the wire direction with magnitudes $h_\mathrm{L}$ and $h_\mathrm{R}$. Later we include a spin-splitting field in the central region captured by the magnetic phase $\Phi_M$.

The discrete spectrum is obtained by solving the equation 
\begin{equation}
    \det[1- S_\mathrm{C}(\varepsilon) S_\mathrm{A}(\varepsilon)] = 0
    \label{eq:matrix_eq1}
\end{equation}
where $S_\mathrm{A}$ is the scattering matrix at the interfaces while $S_\mathrm{C}$ is the scattering matrix for the transmission through the central region. We use the basis $\psi^\mathrm{in} = (c^e_+(0)\, c^e_-(\ell_\mathrm{C})\, c^h_-(0)\, c^h_+(\ell_\mathrm{C}))$ and $\psi^\mathrm{out} = (c^e_-(0)\, c^e_+(\ell_\mathrm{C})\, c^h_+(0)\, c^h_-(\ell_\mathrm{C}))$ such that $S_\mathrm{C} \psi^\mathrm{in} = \psi^\mathrm{out}$ and  $S_\mathrm{A} \psi^\mathrm{out} = \psi^\mathrm{in}$.

For $\varepsilon < \Delta$, we can assume that no normal reflection happens at the interface between the central and lateral regions if the interfaces are clean. In this case, the Andreev reflection matrix takes the form 
\begin{equation}
    S_\mathrm{A} = \begin{pmatrix} 0 & s_\mathrm{A}^{eh} \\ s_\mathrm{A}^{he} & 0 \end{pmatrix} \,.
\end{equation}

    where the submatrices are
\begin{align}
    &s_\mathrm{A}^{eh} = 
    A_\varepsilon e^{+i\tau_z \frac{\phi}{2}}\,,
    &s_\mathrm{A}^{he} =A_\varepsilon e^{-i\tau_z \frac{\phi}{2}}\,,
\end{align}
with
\begin{equation}
    A_\varepsilon = \begin{pmatrix}
    e^{-i \arccos \qty(\frac{\varepsilon + \sigma h_\mathrm{L}}{\Delta})} & 0 \\
    0 & e^{-i \arccos \qty(\frac{\varepsilon + \sigma h_\mathrm{R}}{\Delta})}
     \end{pmatrix}\,.
\end{equation}

For the central region, we assume a general scattering matrix that is block-diagonal in the electron-hole subspaces:
\begin{equation}
    S_\mathrm{C} = 
    \begin{pmatrix}
    s_\mathrm{N}^{ee} & 0 \\
    0 & s_\mathrm{N}^{hh}
    \end{pmatrix}\,.  
\end{equation}

{
Both $S_\mathrm{A}$ and $S_\mathrm{C}$ satisfy particle-hole symmetry 
\begin{equation}
    S(\varepsilon) = \mathcal{P} S(-\varepsilon) \mathcal{P}^\dag =\sigma_y \tau_y S^*(-\varepsilon)  \sigma_y  \tau_y \,,
\end{equation}
and consequently $s_\mathrm{N}^{hh}(\varepsilon) = \mathcal{T} s_\mathrm{N}^{ee}(-\varepsilon) \mathcal{T}^\dag$.
} Using the property 
\begin{equation}
    \det\begin{pmatrix}A & B \\ C & D\end{pmatrix} = \det (A D - A C A^{-1} B)
\end{equation}
we eliminate the particle-hole blocks to simplify Eq.~\eqref{eq:matrix_eq1} to
\begin{equation}
\det (1 - s_\mathrm{A}^{he} s_\mathrm{N}^{ee} s_\mathrm{A}^{eh} s_\mathrm{N}^{hh}) = 0\,.
\label{eq:matrix_eq2}
\end{equation}
To proceed, we need to introduce some specific assumptions on the form of $S_\mathrm{C}$. For a clean system, we can assume a free propagation that, in Andreev approximation, results in a scattering matrix with the  form 
\begin{equation}
s_\mathrm{N}^{ee} = s_\mathrm{N}^{hh} = \exp (i \frac{\pi}{2}\frac{\varepsilon + \sigma h_\mathrm{C}}{E_\mathrm{T}})
\begin{pmatrix}
    0 & 1 \\
    1 & 0 
\end{pmatrix}
\end{equation}
where we defined the Thouless energy $E_\mathrm{T} = \frac{\pi}{2}\frac{\hbar v_F}{\ell_\mathrm{C}}$.
This results in an equation for the bound states
\begin{equation}
    \pi \frac{\varepsilon + \sigma h_\mathrm{C}}{E_T} \pm \phi - \sum_{\mathrm{lat} = L, R} \arccos (\frac{\varepsilon +\sigma h_\mathrm{lat}}{\Delta}) = 2 \pi n\,.
    \label{eq:semiclassical}
\end{equation}
If we take the exchange field in the lateral regions to be equal $h_\mathrm{L} = h_\mathrm{R} = h_\mathrm{lat}$, this simplifies to 
\begin{equation}
    \frac{\pi}{2}\frac{\varepsilon}{E_T} + \sigma \Phi_M \pm \frac{\phi}{2} - \arccos (\frac{\varepsilon + \sigma h_\mathrm{lat}}{\Delta}) = \pi n\,,
\end{equation}
where we defined the magnetic phase $\Phi_M \equiv \frac{\pi}{2}\frac{h_\mathrm{C}}{E_T} = \frac{h_\mathrm{C} \ell_\mathrm{C}}{\hbar v_F}$. In the case of a short central region, we can neglect the spin-independent phase shift in the central region and get the simplified relation 
\begin{equation}
    \varepsilon_{0, \sigma} =  \pm \Delta\cos(\pm \frac{\phi}{2} + \sigma \Phi_M) - \sigma h_\mathrm{lat}
\end{equation}

\begin{figure}[t]
    \centering
    \includegraphics[width=\columnwidth]{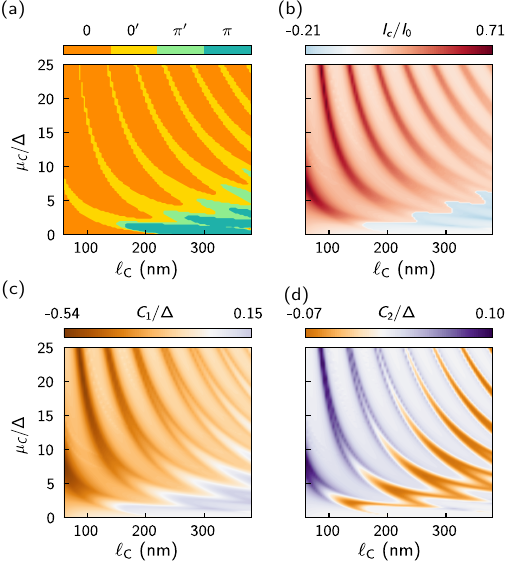}
    \caption{
    \textbf{Dependence of the phase diagram on the length of the central region.}
    The plots show the phase diagram, (a), critical current, (b), and lowest two harmonics, (c) and (d).
    Parameters: $h_\mathrm{lat}=0$, $h_\mathrm{C}=\SI{0.1875}{\milli\eV}$, $\mu_\mathrm{C}=\SI{3.125}{\milli\eV}$ $\mu_\mathrm{lat}=\SI{1.25}{\milli\eV}$, $V_\mathrm{B} = \SI{0.5}{\milli\eV}$.
\label{fig:app:length_dependence}
      }
\end{figure}

In the case of a low-density regime, the conduction-band polarization $h_\mathrm{C}/\mu_\mathrm{C}$ can reach high values and exceed one at the transition to a half-metallic regime. The presence of inhomogeneities in the chemical potential or exchange field causes the breakdown of quasiclassical Andreev solutions that manifests in the hybridization of the solutions of Eq.~\ref{eq:semiclassical} and the opening of sizable gaps in the spectrum \cite{Cayssol_PRB_2004, Cayssol_JMMM_2006}. 

Some features of this simple model can be connected to the results of the non-linearized BdG model displayed in Fig.~\ref{fig:app:length_dependence}. Even in the short superconductor-normal metal-superconductor junction limit ($\ell_\mathrm{C}\ll\xi$), the junction length modulates the magnetic phase acquired in the transport in the normal region. Systems with a longer normal region show an alternation of $0$ and $\pi$ phases together with $0'$ and $\pi'$ regions. The phase diagram shows a series of $\pi$ regions with a $\frac{\ell_\mathrm{C}}{\mu_\mathrm{C}^a}\sim b$ shapes, with $a$ and $b$ constants. This can be understood on the basis of the semiclassical result where the effect of the ferromagnetic insulator in the central region enters the ABSs spectrum through the magnetic phase $\Phi_M = \frac{\pi}{2} h_\mathrm{C} \ell_\mathrm{C} (\frac{2 \mu_\mathrm{C}}{m*})^{-1/2}$. Therefore changing the density can have effects similar to changing the length of the junction.  

\bibliography{references.bib}

\end{document}